\documentclass[12pt,twoside]{article}
\usepackage{fleqn}
\usepackage{espcrc1}
\usepackage[dvips]{graphicx}
\usepackage{amssymb}

\def\pbp{\left\langle\bar{q}q\right\rangle}
\def\chidof{\chi^2 / {\rm d.o.f.}}
\def\MeV{{\rm\  MeV}}

\title{Dynamical lattice QCD thermodynamics with domain wall fermions.}
 
\author{
  George T. Fleming
  \thanks{
    In collaboration with
    P.~Chen,
    N.~Christ,
    A.~Kaehler,
    T.~Klassen,
    C.~Malureanu,
    R.~Mawhinney,
    G.~Siegert
    C.~Sui,
    P.~Vranas,
    L.~Wu,
    Y.~Zhestkov.
    Supported in part by
    DOE grant \# DE-FG02-92ER40699
    and in part by NSF grant \# NSF-PHY96-05199 (PMV).
  }
  \address{Physics Dept., Columbia University, New York NY 10027}
}

\begin{document}

\def\thepage{CU-TP-955}

\maketitle

\begin{abstract}
We present results from simulations of two flavor QCD
thermodynamics at $N_t=4$ with domain wall fermions.
In contrast to other lattice fermion formulations,
domain wall fermions preserve the full
${\rm SU}_L(N_f) \otimes {\rm SU}_R(N_f)$ symmetry of
the continuum at finite lattice spacing (up to terms exponentially small
in an extra parameter).  Just above the phase transition,
we find that the ${\rm U}_A(1)$ symmetry is broken only by a small amount.
We discuss an ongoing calculation to determine the order and properties
of the phase transition using domain wall fermions, since the global symmetries
of the theory are expected to be important here.
\end{abstract}

\section{Introduction}
\label{sec:intro}

Domain wall fermions (DWF) \cite{Kaplan:1992bt} solve the fermion
doubling problem on the lattice through the introduction of
a fictitious internal flavor space.
After lifting the doublers with a Wilson term, the remaining light
fermion respects the ${\rm SU}_L(N_f) \otimes {\rm SU}_R(N_f)$ chiral
symmetry, up to terms exponentially small in size, $L_s$,
of the flavor space.  Since the computing cost is linear in $L_s$
it may be possible to simulate $N_f=2$ QCD on the lattice
with the full flavor and chiral symmetries of the continuum
close to the finite temperature transition using today's supercomputers.

We summarize here the efforts of the Columbia group
to study two flavor QCD thermodynamics with DWF.
Some of these results have also been presented elsewhere.
\cite{Vranas:1998vm,Vranas:1999pm}
The DWF action used in this work is as in \cite{Furman:1995ky} with
the modifications as in \cite{Vranas:1997da}. Some details on the numerical
methods can be found in \cite{Vranas:1998vm}.
For quenched QCD thermodynamic studies with DWF see 
\cite{Fleming:1998cc,Chen:1999kg,Chen:1998xw,Lagae:1999ym},
and with overlap fermions see \cite{Edwards:1999xn}.
For a review on DWF see \cite{Blum:1998ud} and references therein.

\section{The ${\rm U}_A(1)$ symmetry in the deconfined phase}
\label{sec:anomaly}

At low temperatures, the anomalous breaking of the ${\rm U}_A(1)$ symmetry in
QCD has several observable consequences.  For example, it explains the
large mass difference between the $\eta$ and $\eta^\prime$.
At finite temperatures, the role of the chiral anomaly is not as
well understood.  In $N_f=2$ massless QCD, if the deconfinement transition
restores ${\rm U}_A(1)$ as well as ${\rm SU}_L(2) \otimes {\rm SU}_R(2)$
then a first order phase transition is expected.
On the contrary, if ${\rm U}_A(1)$ remains
substantially broken in the deconfinement region, universality arguments
suggest the possibility  of a second order transition.
\cite{Pisarski:1984ms}

Earlier attempts to address this question
\cite{Chandrasekharan:1998yx,Kogut:1998rh}
using staggered fermions could not produce conclusive results because
because of the zero-mode shifts and classical-level ${\rm U}_A(1)$ symmetry
breaking which plague staggered fermions.
At $L_s = \infty$ massless DWF do not break the ${\rm U}_A(1)$ symmetry at the
classical level.
't Hooft first recognized that the ${\rm U}_A(1)$ symmetry is broken by
fermion zero modes which are connected to the topology of the gauge fields.
\cite{'tHooft:1976up}
The DWF Dirac operator can have exact zero modes \cite{Narayanan:1995gw}
and, in an earlier study, \cite{Chen:1998ne} we demonstrated that domain wall
fermions are sensitive to the fermionic zero modes produced by
smooth topological gauge configurations
for fairly small masses and for $L_s \gtrsim 10$.

\pagenumbering{arabic}
\setcounter{page}{2}

\begin{figure}[h]
\hfill
\begin{minipage}{2.5 truein}
  \includegraphics[width=2.5 truein]{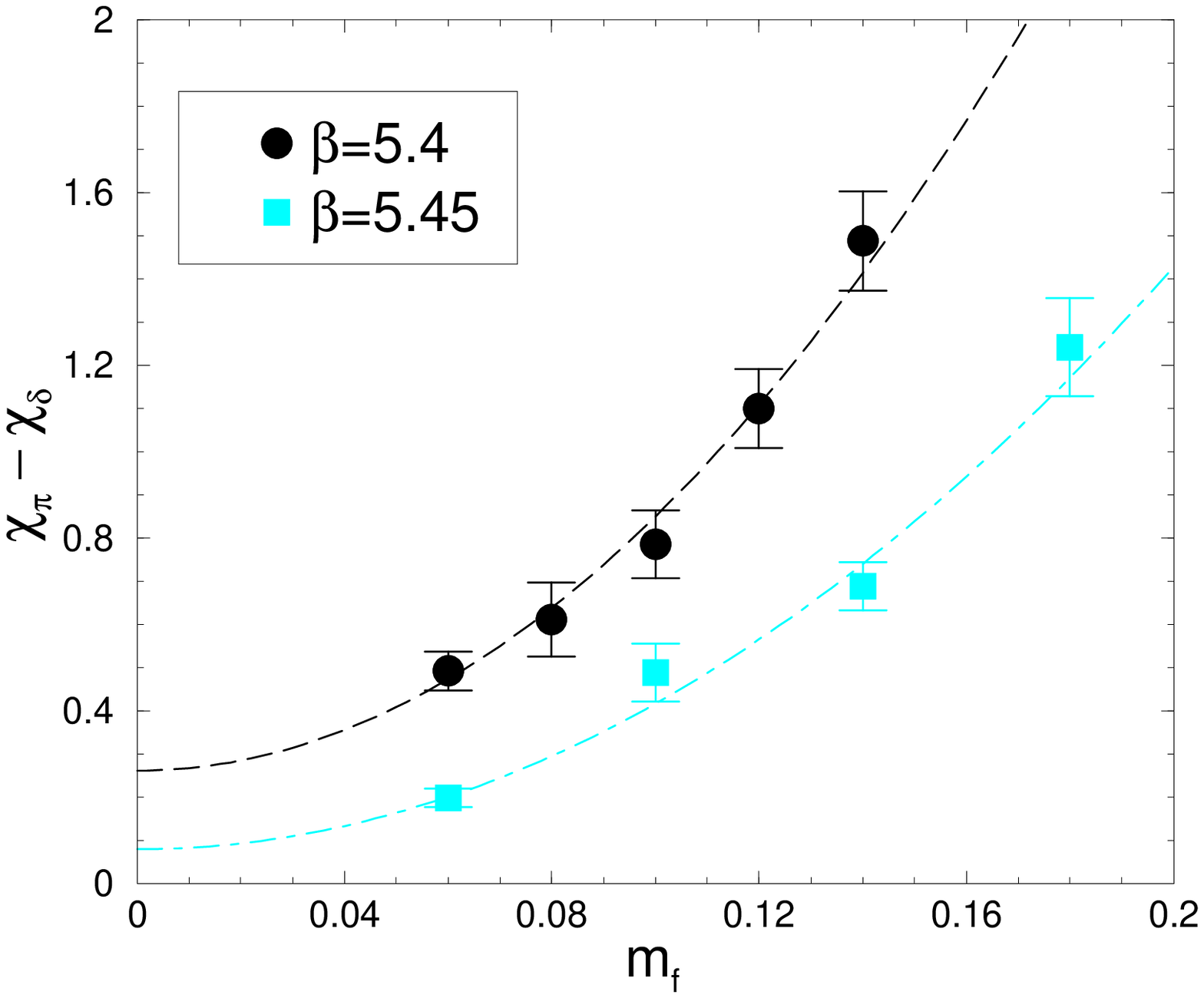}
  \caption{
    Fit of $\chi_\pi - \chi_\delta = c_0 + c_2 m_f^2,$
    vol: $16^3\times 4, m_0=1.9, L_s=16.$
  }
  \label{fig:one}
\end{minipage}
\hfill
\begin{minipage}{2.5 truein}
  \includegraphics[width=2.5 truein]{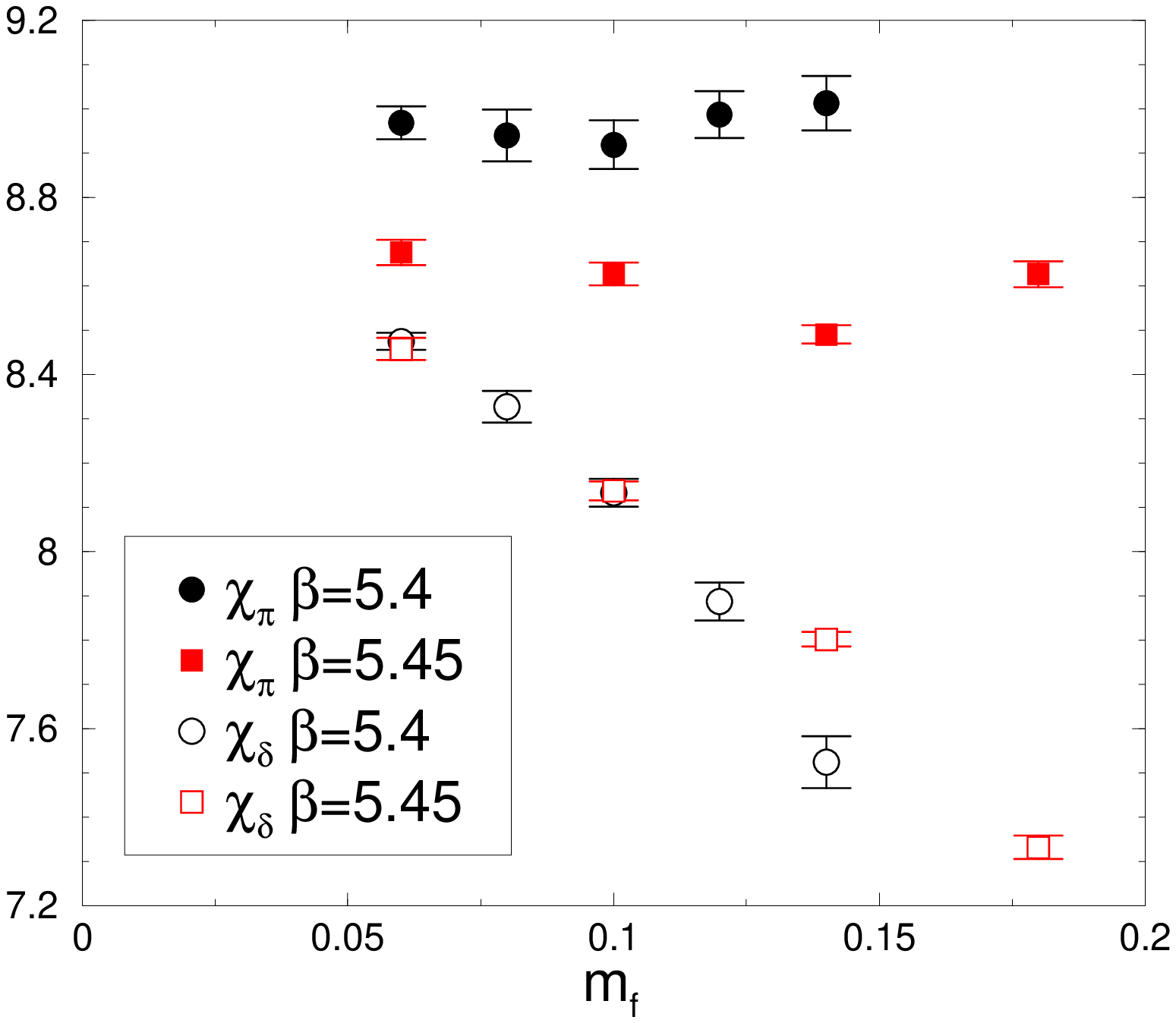}
  \caption{
    $\chi_\pi, \chi_\delta$ vs. $m_f$,
    vol: $16^3\times 4, m_0=1.9, L_s=16.$
  }
  \label{fig:two}
\end{minipage}
\hfill\mbox{}
\end{figure}

To measure anomalous symmetry breaking, we compare the susceptibilities
of the pseudoscalar $\pi$ mesons and scalar $\delta$ mesons.
Their difference, $\chi_\pi - \chi_\delta$, is plotted versus the
dynamical bare quark mass $m_f$ for $\beta=5.4, 5.45$
in figure \ref{fig:one}.  The lattice volume is $16^3\times 4$,
$L_s = 16$, and $\beta_c \approx 5.325$ for $N_t = 4$
(see section \ref{sec:transition}).  The fit ansatz
$\chi_\pi - \chi_\delta = c_0 + c_2 m_f^2$ is also shown.
For $\beta=5.4, c_0=0.26(6)$ and for $\beta=5.45, c_0=0.08(3)$.
For both fits, $\chidof \approx 1$.
We can also conclude that $L_s=16$ is large enough to suppress
the chiral symmetry breaking terms (relative to the bare quark mass),
because the data fits the ansatz without a linear term
within our current statistics.

It is also interesting to compare the magnitude of $\chi_\pi - \chi_\delta$
to the magnitude of the susceptibilities, shown in figure \ref{fig:two}.
We see that although the differences are non-zero by
a statistically significant amount, the scale of the difference is small.
It is an open question as to whether the small size
of the ${\rm U}_A(1)$ symmetry breaking is sufficient to support
a second order phase transition.

\section{In the transition region}
\label{sec:transition}

\begin{figure}[h]
\hfill
\begin{minipage}{2.5 truein}
  \includegraphics[width=2.5 truein]{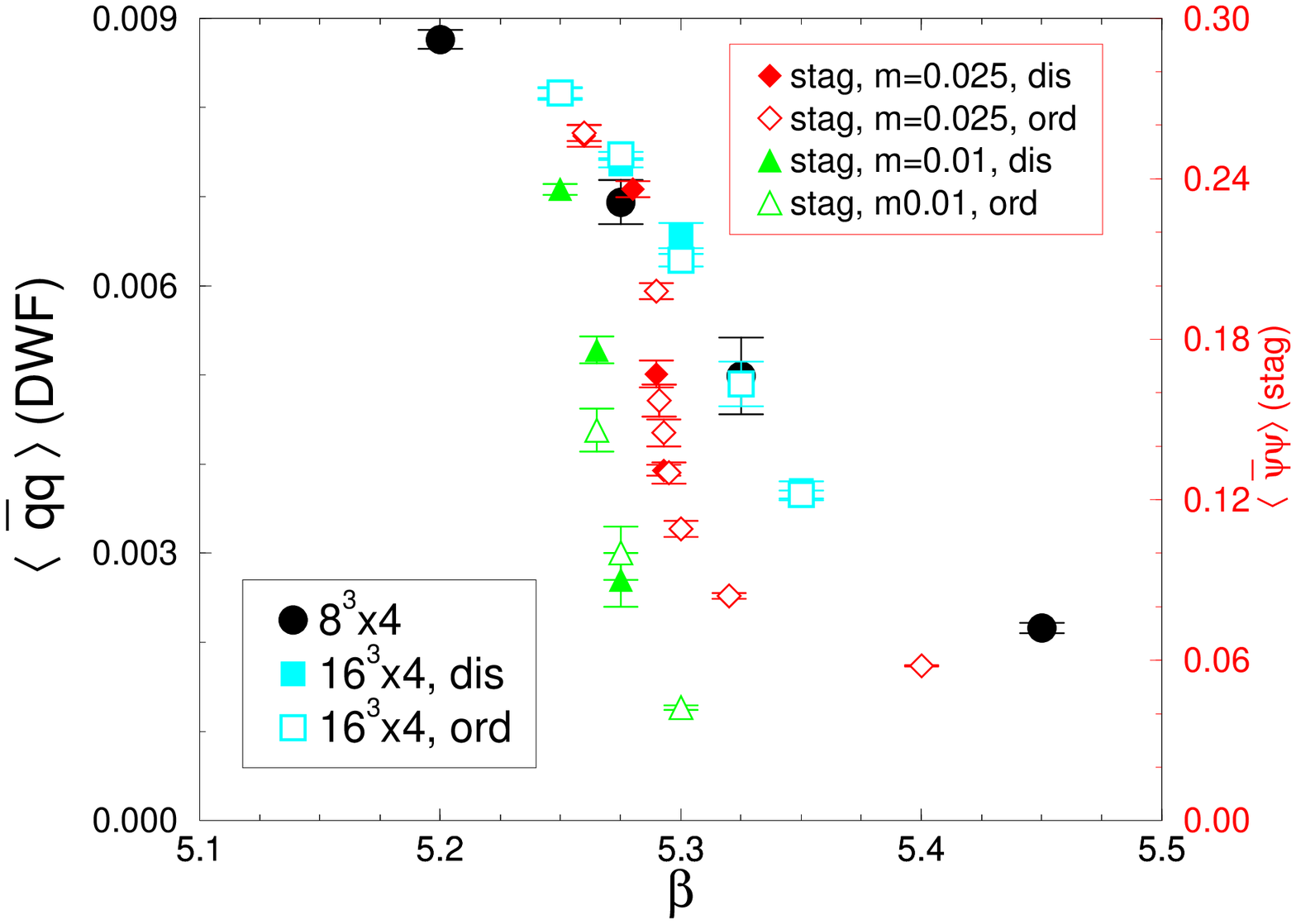}
  \vskip -0.5 truein
  \caption{
    DWF and stag $\pbp$,
    vol: $16^3\times 4, c_1=0, m_0=1.9, L_s=24.$
  }
  \label{fig:three}
\end{minipage}
\hfill
\begin{minipage}{2.5 truein}
  \includegraphics[width=2.5 truein]{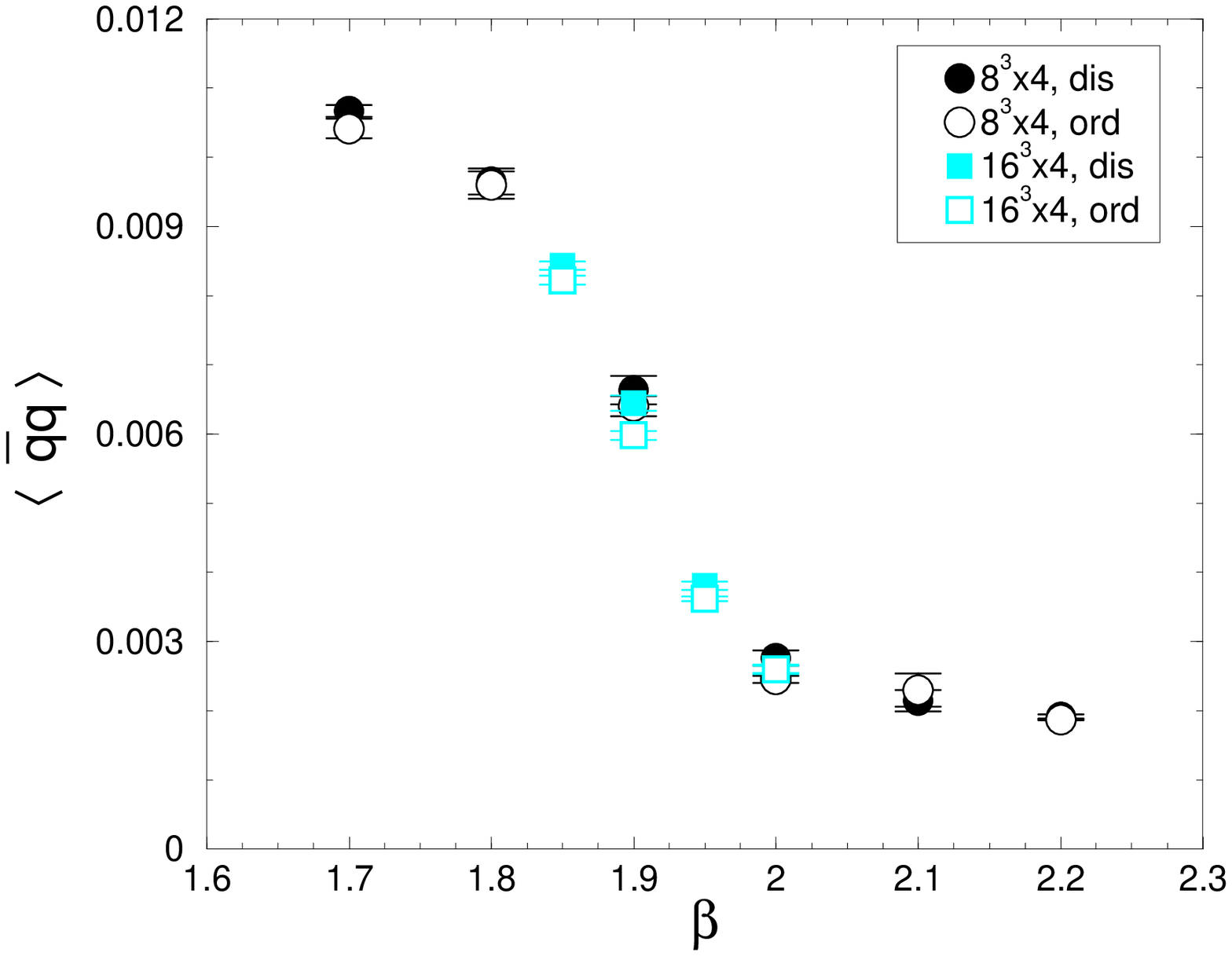}
  \vskip -0.5 truein
  \caption{
    DWF $\pbp$,
    vol: $16^3\times 4, c_1=-0.331, m_0=1.9, L_s=24.$
  }
  \label{fig:four}
\end{minipage}
\hfill\mbox{}
\end{figure}

Based on earlier exploratory studies using smaller $8^3\times 4$ volumes,
\cite{Vranas:1998vm,Vranas:1999pm}
the transition region was localized to $5.2 < \beta_c < 5.4$
for the domain wall height $m_0=1.9$.  To demonstrate critical behavior,
it is important to use large spatial volumes relative to the temporal extent.
We studied this region on $16^3\times 4$ volumes with
$m_f=0.02$ and $L_s=24$.  Initial estimates using $\pbp$ suggested
that chiral symmetry breaking effects would be no larger than 
$\sim 15\%$ throughout the transition region.
The sweep through the region is shown in figure \ref{fig:three}.
The gauge part of the action is the standard Wilson plaquette action,
indicated by $c_1=0$.  For comparison, $N_f=2$ staggered data from
\cite{Vaccarino:1991kn} is shown.

In order to set the scale, we ran an $8^3 \times 32$ simulation
at $\beta=5.325$ corresponding to the middle of the transition region.
\cite{Wu:1999cd}  We found that in lattice units $m_\rho = 1.18(3)$ 
and $m_\pi = 0.654(3)$, yielding a critical temperature $T_c = 163(4) \MeV$
and $m_\pi = 427(11) \MeV$.
The critical temperature is in agreement with results obtained from
other fermion regulators.\cite{Karsch:1999vy}  The pion mass is clearly
too heavy to be able to extract useful information regarding the
order of the transition.  ( By comparison, for staggered fermions
near $T_c$ for $N_t=4$, one pion has a mass of $m_{\pi} \sim 228 \MeV$
and the other two are $m_{\pi_2} \sim 604 \MeV$,
\cite{Vaccarino:1991kn} which might
help explain the different crossover rates between the two regulators.)
Furthermore, a more sophisticated study of residual chiral symmetry
breaking effects indicate that effects of finite $L_s$ are
of the same magnitude as the bare quark mass $m_f$ and suggest
that $L_s \sim 100$ may be necessary to lower the three pion masses
to the physical regime. \cite{Fleming:1999eq}

In an attempt to reduce the chiral symmetry breaking effects due to 
finite $L_s$, we improved the gauge action by adding a $1\times 2$
rectangular plaquette to the standard Wilson gauge action,
with the choice of coefficient, $c_1=-0.331$, suggested by Iwasaki.
\cite{Iwasaki:1983ck}
An initial study of the quenched hadron spectrum
at couplings corresponding to the quenched $N_t=4$ QCD transition region
indicated a significant reduction in finite $L_s$ effects.
\cite{Wu:1999cd}  These results prompted a study of the $N_f=2$ QCD
transition region using Iwasaki improved gauge action and DWF.
The results are presented in figure \ref{fig:four}.

Again, we set the scale simulating meson masses on an $8^3 \times 32$ lattice
at $\beta=1.9$, corresponding to the middle of the transition region.
\cite{Wu:1999cd}
In lattice units, we found $m_\rho = 1.163(21)$ and $m_\pi = 0.604(3)$,
which gives $T_c = 166(3) \MeV$ and $m_\pi = 400(7) \MeV$.  
The critical temperature is in agreement with the
standard Wilson gauge action results. However, the pion mass
is not significantly lighter.

\section{Conclusions}
\label{sec:conclusions}

We measured the difference $\chi_\pi - \chi_\rho$
just above the $N_t=4$ finite temperature
transition and found to be non-zero by a statistically significant amount.
However, the magnitude of the difference 
is much smaller than the susceptibilities themselves.
Thus, while the ${\rm U}_A(1)$ is broken in the symmetric
phase it is not clear what effect this may have on the order
of the transition.

We studied the transition region using two different gauge actions.
In both cases, the transition
appears to be a smooth crossover, probably due to larger than
expected chiral symmetry breaking effects of finite $L_s$.
We then ran zero temperature simulations of meson masses
at couplings corresponding to the middle of the crossover region
to set the scale.
In each case, the critical temperatures and pion masses we extracted
are comparable to other lattice regulators.
Alternate techniques for reducing chiral symmetry breaking at finite $L_s$
are required to reduce the pion masses to physical values
as simply repeating the study at larger $L_s$ would take
several years using current supercomputers.

All calculations were done on the $400$ Gflops QCDSP machine at
Columbia University.


\end{document}